
\input phyzzx
\Pubnum={SLAC--PUB--5810\cr
TUM-TH-144/92\cr
SNUTP 92-26}
\date{April 1992}
\pubtype{T/AS}
\titlepage
\title{AXINO MASS\doeack}
\author{E. J. Chun}
\address{Physik Department, Technische Universit\"at M\"unchen\break
D-8046 Garching, Germany}
\author{Jihn E. Kim}
\address{Center for Theoretical Physics and Department of Physics\break
Seoul National University, Seoul 151-742, Korea}
\andauthor{H. P. Nilles\footnote{\#}{On leave of absence from
Physik Department, Technische  Universit\"at M\"unchen
 and Max-Planck Institut
f\"ur Physik, P.O. Box 40 12 12, D-8000 M\"unchen. Work partially
supported by Deutsche Forschungsgemeinschaft.}}
\SLAC
\abstract
The mass of the axino is computed in realistic supersymmetric extensions
of the standard model. It is found to be strongly model dependent
and can be as small as a few keV but also as large as the
gravitino mass. Estimates of this mass can only be believed once
a careful analysis of the scalar potential has been performed.
\endpage

While the standard model of electroweak interactions is in complete
agreement with all experimental findings, still much effort has been
devoted to construct various generalizations.
Supersymmetric extensions have attracted much attention in the last
decade\rlap.\REF\Nilles{H. P. Nilles \journal
 Physics Reports& 110 (84) 1}\refend\
 Realistic models of this type consider local supersymmetry
(supergravity), spontaneously broken in a so-called hidden sector
at a mass scale of $M_S\approx 10^{11}$GeV. The induced
SUSY-breaking mass scale in the observable sector is  given
by a value of order of the gravitino mass
$m_{3/2}\sim M_S^2/M_P$, where $M_P$ denotes the Planck mass.
The breakdown scale of the weak interaction gauge symmetry
$SU(2)\times U(1)$ is then closely related to the scale of
SUSY-breakdown in the observable sector, thus explaining the
smallness of $M_W$ compared to $M_P$, once we understand the
mechanism for the breakdown of supersymmetry.

In the minimal supersymmetric extension of the standard model,
however, there is one dimensionful parameter which is not proportional
to $m_{3/2}$, the Higgs mass term $\mu H\bar H$ in the superpotential.
Such a term is allowed by supersymmetry and it remains to be understood
why also $\mu$ should take values in the desired energy range.
One suggestion to achieve this starts with the consideration of
the axionic generalization of the standard model\rlap.\REF\Kim{
Jihn E. Kim \journal Physics Reports& 150 (87) 1}\refend\
The  decay
constant $F_a$ of the invisible axion should lie in the range given
by $M_S$ and it was argued that such a coincidence cannot be accidental.
In fact, one can easily construct models, in which $\mu$ is generated
dynamically to be in the range of the gravitino mass\rlap.\REF\KN{
J. E. Kim and H. P. Nilles \journal Phys. Lett.& B138 (84) 150}\refend\

Of course, the prime motivation to consider the axionic generalization of
the standard model is the quest for a natural solution of the
strong CP-problem. A supersymmetric realization of this mechanism
is most easily achieved in
a model with several Higgs supermultiplets\rlap.\REF\NR{H. P.
Nilles and S. Raby \journal Nucl. Phys.& B198 (82) 102}\refend\
Instead of just the pseudoscalar axion, such a model now possesses a full
axion supermultiplet. This contains the {\it axino}, the fermionic
partner of the axion, as well as the {\it saxino},
the
scalar partner of
the axino. Although these particles are very weakly interacting,
they might nonetheless lead to important astrophysical and
cosmological consequences.
The stability of stars and the observed energy density of the universe,
as is well known, restrict the decay constant of the invisible axion
to a small window\rlap.\REF\Raffelt{G. G. Raffelt \journal
Physics Reports& 198 (90) 1}\refend\
Furthermore, a very light saxino might give rise
to new long range interactions which are incompatible with present
observations. In a model of broken supersymmetry, however, one would
usually expect the saxino to receive a mass of order of the gravitino
in the 100 GeV to TeV range, avoiding such unpleasant phenomena
as a fifth force.

The possible consequences of the presence of an axino have not been
considered in detail until recently\rlap.\REF\RTW{K. Rajagopal,
M. S. Turner and F. Wilczek \journal Nucl. Phys. & B358 (91) 447
}\refend\
Given the weakness of its interactions, of course, we would rather
expect to find only indirect manifestations of the existence of
such a particle. In fact, up to now only the effects of the
(possibly) stable axino on the total energy density of the universe
have been studied. A stable axino in a certain mass range could lead
to overcritical energy density, and the corresponding axion models
are therefore ruled out. If, on the other hand, the axino has a mass of
a few keV, it is itself an interesting candidate\refmark\RTW\
for a source of dark
matter. In fact, such a particle is up to now the only well motivated
candidate for so-called {\it warm} dark matter. It still remains to be
seen, however, whether warm dark matter can lead to a satisfactory
cosmological model including questions about large scale
structure formation.

In any case, the existence of a (light) axino might have important
consequences in any supersymmetric extension of the standard
model. Many of these models contain discrete symmetries
(R-parity in the simplest case) that allow only pair production of
the new supersymmetric particles. In these cases, there exists a
lightest supersymmetric particle (called LSP) which is stable. In
the minimal model one usually considers such weakly interacting
massive particles (WIMPS, an example can be found in the photino) as
a possible source of {\it cold} dark matter. In the presence of
an axion supermultiplet, it could very well happen, that the axino is the
LSP and thus render the WIMP unstable, at least on cosmological
time scales.

While other properties of the axino seem not to be so important for
our discussion, its mass is a crucial parameter and
a careful analysis is required. We shall see in the following that this
value is strongly model dependent. Before we discuss these questions
in detail, let us remark, however, that,
in general supergravity models,
there are some {\it natural}
values such a mass can have. One of
them could be the mass of the gravitino that sets the scale of
SUSY breakdown in the observable sector. But this is not the only
possibility. The axino could very well be much lighter. In fact,
models based on supergravity contain a very small dimensionless
parameter
$$\eta={M_S\over{M_P}}\approx 10^{-8}.\eqno(1)$$
The gravitino mass is then given by
$m_{3/2}\sim \eta^2 M_P$
and the natural values for the mass of the axino at the tree level are
given by
$$m_{\tilde a}\sim \eta^k M_P\eqno(2)$$
with $k\geq 2$.

It is interesting to observe, that in the case of $k=3$ this mass is
in the region of 1 to 10 keV, leading to a critical mass density of the
universe. In models of global supersymmetry
one obtains similar estimates. Here $\eta\sim M_W/F_a$ where $M_W$
denotes the scale of weak interaction
breakdown\rlap.\refmark\RTW\ These values
coincide, since $M_S$ and $F_a$ are so close to each other.

The task of determining the mass of the axino in a given model now
boils down to the question about the power $k$ appearing in (2).
For large $k$, of course, also radiative corrections to $m_{\tilde a}$
have to be taken into account.

Let us start our discussion in the framework of globally supersymmetric
models. Although the construction of supersymmetric generalizations
nowadays exclusively considers locally supersymmetric (supergravity)
models, we can still learn a lot from the simpler models based on global
SUSY. In the present example we can see quite easily, why it makes
sense to consider the possibility of a very small axino mass.
In the case of unbroken supersymmetry, the whole axion supermultiplet
will remain degenerate at the mass given by the anomaly, which
we shall neglect in the following. Thus the mass of the axino and the
saxino have to be proportional to the scale of SUSY breakdown
represented by the vacuum expectation value of an auxiliary field $F_G$
(this is the auxiliary field of the goldstino multiplet)\footnote{
\dag}{$F_G$ is in general a combination of the auxiliary
fields of gauge and chiral supermultiplets.}.
The mass splitting of the chiral supermultiplet
is  determined by the coupling of
its members\refmark\Nilles\  to $F_G$. The axion is protected by
a symmetry and does not receive a mass in the presence of SUSY breakdown.
The scalar saxino couples in general to $F_G$ and will thus obtain  a
mass of the order $g\VEV{F_G}$, where $g$ is the coupling to the
goldstino multiplet\footnote{*}{Actually, in many models based
on global supersymmetry this coupling can be very small and
even vanish at tree level. These vanishing scalar masses, however,
are the reason, why globally supersymmetric models do not lead to a
realistic generalization of the standard model. We shall come
back to this point later.}.
This is the reason why one usually assumes the
saxino to be heavy. In the case of the axino the situation
is similar, but different. Again its mass is determined by the coupling
to $F_G$, but the auxiliary field has canonical dimension two.
A mass term for the axino $\tilde a\tilde a F_G$ is of dimension five
and there are {\it no renormalizable contributions to the mass of the
axino}. In  a model with an (invisible) axion we have as additional
dimensionful parameter the axion decay constant $F_a$ of order of
$10^{11}$ GeV and we therefore expect a small axino mass
\REF\TW{K. Tamvakis and D. Wyler \journal Phys. Lett.& B112 (82) 451}
$m_{\tilde a}\sim {F_G\over{F_a}}$ as was demonstrated in ref. \TW.

It remains to be seen, how these results generalize once we consider
models based on supergravity. The reason why one nowadays primarily
considers these models is the fact  that in models based on
spontaneously broken global SUSY a universal mass shift for the scalar
partners of quarks and leptons is not possible. We have mentioned that
already in connection with the discussion of
the mass of the saxino. This fact holds for a large class of models
and can be succinctly summarized by the value of
$STr M^2$, the supertrace of the square of the mass matrix. These
results suggest that in realistic models the masses of the scalars,
and thus  also the mass of the saxino, are pushed up to a value
beyond the reach of present experiments. In the case of the axino
such a general statement cannot be made. The authors of ref. \RTW\ assume
(in order to avoid the murky depths of supergravity theory as they say),
that the globally supersymmetric results carry over to the supergravity
case. We shall see in the following that, in general, such an assumption
is not necessarily correct. A similar conclusion has been obtained by
Goto and Yamaguchi\rlap.\REF\GY{T. Goto and M. Yamaguchi, Tohoku
University preprint TU-378 (1991)}\refend\
Their result seems to to imply, however,
that a small mass of
the axino requires a special form of the kinetic terms.
We analyze this issue in a more general way and see that, independent
of the choice of the kinetic terms, small (and also large) axino masses
are possible, dependent on other properties of the theory.
We also investigate the question of the axino mass in those models
that might be found as the low energy limit of string theory.

The scalar sector of a supergravity theory is completely specified
by the K\"ahler potential $G(\Phi^j,\Phi_j^*)$ where $\Phi$ collectively
denotes the chiral superfields. The scalar kinetic terms are given by
the second derivative $G^i_j=\partial^2G/\partial\Phi^j\partial
\Phi_i^*$ and one often splits
$G(\Phi,\Phi^*)=K(\Phi,\Phi^*)+\log\vert W(\Phi)\vert^2$
where the superpotential $W(\Phi)$ is a holomorphic function of
$\Phi$.
The scalar potential is given by\refmark\Nilles
$$V =-\exp G\left[3-G_i(G^{-1})^i_j G^j\right].\eqno(3)$$
We are interested in the mass spectrum of the theory once supersymmetry is
broken spontaneously, which leads to a nontrivial value of the
gravitino mass $m_{3/2}^2=\exp(G)$\footnote{\dag}{We assume
vanishing vacuum energy, thus
$\VEV{V}=\VEV{V_i}=0$ at the minimum.}.
Masses of the scalar particles can then be read off from the second
derivative of the potential at the minimum. For the fermions we
obtain
$$M_{ij}=\exp(G/2)\left[ G_{ij}+{1\over 3}G_iG_j
-G_k(G^{-1})^k_l G^l_{ij}\right],\eqno(4)$$
where we have removed the contribution to the mass of the gravitino.
We also have to respect the constraint from the anomalous $U(1)$-symmetry
$$\sum_i \left[ q_i\Phi^i G_i-q_i\Phi_i^* G^i\right] =0,\eqno(5)$$
where $q_i$ is the $U(1)_{PQ}$-charge of $\Phi_i$.

Fields in the observable (hidden) sector shall be denoted by
$y_i$ ($z_i$), respectively, and we shall assume the superpotential to
split: $W(\Phi_i)=h(z_i)+g(y_i)$. In our examples we use the
well known case $h(z)=m^2(z+\beta)$ for simplicity. Let us start
our discussion with a special choice
$G_i^j=\delta_i^j$, usually referred to as minimal kinetic terms.
The scalar potential then reads
$$V=\exp(K/M^2)\left[\vert h_z+{z^*W\over M^2}\vert^2+
\vert g_i+{y_i^*W\over M^2}\vert^2
-{3\vert W\vert^2\over M^2}\right].\eqno(6)$$
Within this framework Goto and Yamaguchi\refmark\GY\
 have argued that the axino
mass is as large as the gravitino mass. Let us see how this works
using their superpotential
$g_1=\lambda(AB-f^2)Y,$
where $f$ is a constant and $A$, $B$ and $Y$ are fields. Minimizing
(6) we find the following (approximate) vacuum expectation
values (vevs): $A=B\approx f$ and $Y\approx m^2/M$, while the $z$ vev
remains undisturbed\footnote{*}{Observe that in the case of
global supersymmetry the minimum is found at
$\VEV{Y}=0$.}.
Actual values for $f$ and $m$ should lie in the
range of $10^{11}$ GeV. Denoting the fermions in the chiral
supermultiplets by $\chi^i$, the axino is found to be the
linear combination
$\chi^a=(\chi^A-\chi^B)/\sqrt 2$ and it receives a mass of order
of the gravitino mass $m_{\tilde a}\sim m_{3/2}\sim m^2/M$.

Is this now a generic property of models with minimal kinetic terms?
We shall see that the answer is no by inspecting a second example
with superpotential $g_2=\lambda(AB-X^2)Y+{\lambda^\prime\over 3}
(X-f)^3$ including a new singlet chiral superfield $X$. Minimization of
the potential now becomes more complicated since the condition
$\VEV{V}=0$ leads to a shift in the vev of the hidden sector fields.
The easiest way to discuss the potential is by expanding it in powers
of $m/M$. To lowest order one obtains the globally supersymmetric
result for the observable sector.
In each order one then has to adjust the vacuum energy to zero,
and in the present example the inclusion of the terms of order
$m^2/M^2$ require a shift of $\VEV{z}$. In the previous example
it was sufficient to just consider the expansion up to first order.
One still obtains vevs of $A$, $B$ similar to those of the
previous example, but the presence of the field $X$ has important
consequences on the axino mass; in fact here one obtains
$m_{\tilde a}\sim m^3/M^2$.

This example shows that the mass of the axino depends strongly on the
model and the special form of the superpotential. It also shows that in
models with minimal kinetic terms the mass of the axino not
necessarily needs to be as large as the gravitino mass, contrary to the
impression given in ref. \GY.
In particular, masses of the axino in the range of a few keV can be
obtained also in this framework.

Let us next consider those supergravity models that have a structure
similar to those that appear in the low-energy limit of string
theories. The K\"ahler potential is given by\REF\Witten{E. Witten
\journal Phys. Lett.& B155 (85) 151}\refend
$$K=-\log(S+S^*)-3\log(T+T^*-C^iC_i^*),\eqno(7)$$
where $S$ denotes the dilaton superfield, $T$ represents the moduli
and $C_i$ the  matter fields.
We shall assume the superpotential of the form
$W=W(S)+W(C_i)$, postponing a discussion of the implications of moduli
dependence. The term $W(S)$ is assumed to appear as a result of
gaugino condensation in the underlying string model, and is crucial
for the process of supersymmetry breakdown. For a review and details
\REF\Gaugino{H. P. Nilles \journal Int. Journ. of Mod. Phys.&
A5 (90) 4199}
see ref. \Gaugino.
The scalar potential of
the theory defined in this way
$$V=\exp(G)\left[\vert G_S\vert^2(S+S^*)+\vert W_i\vert^2
\right]\eqno(8)$$
is positive with a minimum at
$\VEV{V}=0$; the dilaton adjusts its vev to cancel any possible
contribution to the vacuum energy. Supersymmetry is broken spontaneously
through a nontrivial vev of the auxiliary field of the dilaton
supermultiplet $F_T\sim \exp(G) G_T$, while $F_S=0$. The only problem
with the potential is the fact that the vevs of the moduli are not
determined and thus the vacuum is highly degenerate. Let us
nonetheless discuss this simplified example first.
The minimum of (8) is found at $\VEV{G_S}=\VEV{W_i}=0$, where
$W_i=\partial W/\partial C_i=0$ coincides with the solution
obtained in the case of global supersymmetry, independent of the
special form of the superpotential. In our case we require nontrivial
vevs $\VEV{C_i}=v_i$ for at least one of the charged scalar fields.
The axino is then given by $\tilde a=\sum_i({q_iv_i\over v})\chi^i$,
where $v=\sqrt{\sum_i q_i^2v_i^2}$ should take a value of order of $m$
in the $10^{11}$ GeV range.
The goldstone fermion is given by $\eta\sim G_T\chi^T+G_i\chi^i$, with
$G_T=-3/\Delta$, $G_i=3v_i/\Delta$ and $\Delta=T+T^*-C^iC_i^*$. One thus
obtains $\sum q_i v_iG_i=0$ for the axino to be orthogonal to the
goldstino. Fermion masses can now be computed according to (4) in a
straightforward manner. This gives e.g.
$M_{TT}\sim G_{TT}+{1\over 3}G_TG_T+{2\over\Delta}G_T=0$, since
$G_{TT}=3/\Delta^2$ and $G_T=-3/\Delta$. Also the terms mixing $T$- and
$i$-components vanish as well as
$M_{aj}=\sum_i ({q_i v_i \over v}) M_{ij}$
because of the constraint (5).
Thus all these fermions including the axino remain {\it massless}.

One could have expected such a result from the outset
 because of the fact that models
with kinetic terms of the structure (7) are very closely related to
globally supersymmetric models. We have confirmed that above finding
$W_i=0$ at the minimum, the globally supersymmetric solution.
 Thus one might obtain a light
axino in a natural way\rlap.\refmark\GY\ But this is
probably not the whole story. The other fermions remain massless as
well and, more importantly, also the scalar particles like the
saxino remain massless at tree level.
At the present stage of the discussion we can conclude that
this model not only shares the desirable features of globally
supersymmetric models but also the more problematic ones.
Observe that in the models based on minimal kinetic terms the scalar
particles and thus also the saxino received a large mass of order of
$m_{3/2}$.

Again the question arises whether in models with
K\"ahler potential as in (7) one always obtains a light axino.
Unfortunately this question cannot yet be answered definitely.
One way to proceed is to compute radiative corrections
and see how axino and saxino masses
 are shifted\rlap.\refmark\GY\ We would like
to argue, however, that this is not necessarily the correct way to
attack this problem. After all the potential given in (8) has still
a large vacuum degeneracy and many massless scalars and thus is
unstable under small changes of the parameters. In fact,
naively including radiative corrections might destabilize the
potential in such a way that it becomes unbounded from
below. As long as
we do not know the correct position of the minimum we can
not really be sure that our estimate of the axino mass is
reliable.
This can be demonstrated quite easily in the framework of
explicit models.
We have seen this in our discussion of the models with
minimal kinetic terms comparing those with superpotentials
$g_1$ and $g_2$. Although there the potential is less unstable
the actual value of the axino mass
strongly depends on details of the potential.
Similar things will happen also in models with nonminimal
kinetic terms.

In addition we know that the potential as given in (8) is incomplete.
In a first step one should include the moduli-dependent
contributions to the superpotential.
Unfortunately the incorporation of such a dependence in $W$ leads to
enormous complications. The potential is no longer positive
definite and nobody succeeded yet to find a satisfactory minimum
with broken supersymmetry and a vanishing cosmological constant.
As long as such a result is missing, any reliable computation of the axino
mass in such models is impossible. Unfortunately this is also true
in those models with a composite axino that constituted our prime
motivation to study these questions in detail\rlap.\REFS\KNi{J. E. Kim
and H. P. Nilles \journal Phys. Lett.& B263 (91) 79}
\REFSCON\CKN{E. J. Chun, J. E. Kim and H. P. Nilles
\journal Nucl. Phys.& B370 (92) 105}\refsend
This does not mean that the axino cannot be light. In fact our discussion
of the models with minimal kinetic terms has demonstrated that light
axinos could actually exist. We want to stress here that any
estimate of axino masses is unreliable as long as a detailed calculation
of the underlying potential has not been performed.

\ack

The work of E. J. Chun is supported by a KOSEF-fellowship.
J. E. Kim is supported by KOSEF through the Center for
Theoretical  Physics at Seoul National University.

\refout

\end